# On the role of theory and modeling in neuroscience

Daniel Levenstein[1,19], Veronica A. Alvarez[2], Asohan Amarasingham[14], Habiba Azab[3], Zhe S. Chen[18], Richard C. Gerkin[4], Andrea Hasenstaub[5], Ramakrishnan Iyer[6], Renaud B. Jolivet[7], Sarah Marzen[12], Joseph D. Monaco[8], Astrid A. Prinz[13], Salma Quraishi, Fidel Santamaria[9], Sabyasachi Shivkumar[15], Matthew F. Singh[10], Roger Traub[11], Horacio G. Rotstein[16,*], Farzan Nadim[16,*], A. David Redish[17,*]

[1] Montreal Neurological Institute, McGill University, Montreal, QC, Canada
[2] Laboratory on Neurobiology of Compulsive Behaviors, National Institute on Alcohol Abuse and Alcoholism, NIH. Bethesda, MD
[3] Department of Neuroscience, Center for Magnetic Resonance Research, University of Minnesota, Minneapolis, MN
[4] School of Life Sciences, Arizona State University, Tempe, AZ
[5] Department of Otolaryngology - Head and Neck Surgery, University of California San Francisco, San Francisco, CA
[6] Allen Institute for Brain Science, Seattle, WA
[7] Maastricht Centre for Systems Biology (MaCSBio), Maastricht University, Maastricht, Netherlands
[8] Department of Biomedical Engineering, Johns Hopkins University School of Medicine, Baltimore, MD
[9] Department of Biology, University of Texas at San Antonio, San Antonio, TX
[10] Department of Psychological & Brain Sciences, Department of Electrical & Systems Engineering, Washington University in St. Louis, St. Louis, MO
[11] IBM T.J. Watson Research Center, AI Foundations, Yorktown Heights, NY
[12] W. M. Keck Science Department, Pitzer, Scripps, and Claremont McKenna Colleges, Claremont, CA
[13] Department of Biology, Emory University, Atlanta, GA
[14] Department of Mathematics, City College of New York; Departments of Biology, Computer Science, and Psychology, The Graduate Center; City University of New York
[15] Brain and Cognitive Sciences, University of Rochester, Rochester, New York, United States of America
[16] Federated Department of Biological Sciences, New Jersey Institute of Technology and Rutgers University & Institute for Brain and Neuroscience Research, New Jersey Institute of Technology, Newark, NJ
[17] Department of Neuroscience, University of Minnesota, Minneapolis MN
[18] Department of Psychiatry, Neuroscience & Physiology, New York University School of Medicine
[19] Mila - Quebec AI Institute, Montreal, QC, Canada
*Co-senior author

## Abstract

In recent years, the field of neuroscience has gone through rapid experimental advances and a significant increase in the use of quantitative and computational methods. This growth has created a need for clearer analyses of the theory and modeling approaches used in the field. This issue is particularly complex in neuroscience because the field studies phenomena that cross a wide range of scales and often require consideration at varying degrees of abstraction, from precise biophysical interactions to the computations they implement. We argue that a pragmatic perspective of science, in which descriptive, mechanistic, and normative models and theories each play a distinct role in defining and bridging levels of abstraction will facilitate neuroscientific practice. This analysis leads to methodological suggestions, including selecting a level of abstraction that is appropriate for a given problem, identifying transfer functions to connect models and data, and the use of models themselves as a form of experiment.

Correspondence: daniel.levenstein@mcgill.ca, horacio@njit.edu, farzan@njit.edu, redish@umn.edu

## Introduction

Recent technological advances in neuroscience have prompted the growth of new experimental approaches and subfields that investigate phenomena from single neurons to social behavior. However, rapid growth has also revealed a need to develop new theoretical frameworks (Phillips 2015) that integrate the growing quantities of data and to establish relationships between their underlying processes. While neuroscience has a strong history of interactions between experimental and theoretical approaches (Marr 1991; Hodgkin and Huxley 1952; O'Keefe and Nadel 1978), there is still disagreement as to the nature of theory and its role in neuroscience, including how it should be developed, used, and evaluated by the community (Goldstein 2018; Bialek 2018).

We argue that an idealized view of scientific progress, in which science is a problem-solving enterprise that strives to explain phenomena, is well-suited to inform scientific practice. In neuroscience, the phenomena of interest are those that pertain to neurons, the nervous system, and its contribution to cognition and behavior. Because these phenomena span a wide range of spatiotemporal scales, their explanations often require a "multi-level" approach that combines data from dramatically different modalities. **Descriptive**, **mechanistic**, and **normative explanations** each play distinct roles in building a multi-level account of neural phenomena — descriptive explanations delineate an abstract characterization of a phenomenon, while mechanistic and normative explanations bridge abstractions of different levels. Collectively, these operations unify scientific theories across disparate experimental approaches and fields. We show how this view facilitates the bidirectional interaction between theory and experimentation as well as theory development.

## What is a theory and what is it good for?

Theories are the primary tools by which scientists make sense of observations and make predictions. Given this central role, it is surprising how little methodological attention is given in scientific training to the general nature of theories. Traditional descriptions of science tend to be based on the processes of theory identification and falsification, in which theories are proposed as universal truths about the world, tested, provisionally accepted if found to be compatible with experimental data, and rejected when found to be incompatible (Popper 1959). According to these traditional descriptions, when theories are incompatible with experimental data, the conceptual framework on which they are based is called into question and a new framework is found that can better account for the data (Popper 1959; Kuhn 2012; Lakatos 1980). However, historical, philosophical, and sociological analyses argue that these views do not account for how theory is used in practice (Lakatos 1980; Firestein 2015; Godfrey-Smith 2003; Feyerabend 1993; Ben-Ari 2011; Kaiser 2014; Laplane et al. 2019). For example, theories are rarely, if ever, decisively testable, scientists can have a variety of attitudes towards a theory rather than to simply accept or reject it (Lakatos 1978; van Fraassen 1980; Mermin 1989; Ben-Ari 2011; Kaiser 2014), and although new discoveries can provide answers to open questions, the new questions they prompt may be more consequential (Firestein 2012).

### *A pragmatic view: science as problem-solving*

We propose that a pragmatic view of the scientific enterprise (James 1907; Ben-Ari 2011; Laudan 1978; Douglas 2014) is better suited to inform scientific practice. In this view, science is a process through which we solve empirical problems and answer questions about observable phenomena (Laudan 1978; Douglas 2014; Firestein 2015; A. D. Redish et al. 2018; Haig 1987; Nickles 1981). Empirical problems can range from matters of basic scientific interest (for example, "How does the brain process visual signals?" or "How does an animal select between alternative choices?"), to those with more obvious applications (such as "Which brain functions are disrupted in schizophrenia?"). Like any other problem, a scientific problem can be seen as a search to achieve a desired goal, which is specified by the statement of the problem (Newell and Simon 1972). However, scientific problems are often ill-defined (Bechtel and Richardson 2010), in part because the search space and solution criteria are not always explicitly stated and in part because they evolve with additional discoveries (Firestein 2012). For example, the discovery of multiple memory and decision-making systems raises further questions of how those systems interact (Balleine and Dickinson 1998; O'Keefe and Nadel 1978; Daw, Niv, and Dayan 2005; A. David Redish 2013; Scoville and Milner 1957; Squire 1987; Nadel 1994; Schacter 2001), while the question "How does the pineal gland generate consciousness?" (Descartes 1637) is now considered outdated. Further, scientific problems are never definitively solved, but are only deemed "adequately solved" by a research community. What is seen as an adequate solution in one socio-historical context may not be in another — as new data become available, standards change, or alternative solutions are presented. While a continuously evolving landscape of problems and proposed solutions might seem to counter a notion of progress in science, scientific theories have been used to explain and control progressively more phenomena over the course of the scientific record (Laudan 1978; Douglas 2014). According to the pragmatic view, this progress results from community-maintained standards of explanation, under an overarching drive to better predict and control natural phenomena of potential relevance to society (Hacking 1983; Douglas 2014).

We can thus define a scientific explanation (Hempel and Oppenheim 1948; Woodward 2019) as a proposed solution to an empirical problem, and scientific theories to be the ideas we use to form explanations. Where traditional views have tried to specify the form theories take, the pragmatic view sees theory structure as closely tied to its function and context. As a result, a theory can include a wide and complex range of structural elements, including those that are not formalized (Winther 2021). While theories may be spelled out in the scientific literature, they are more often used implicitly in the explanation of phenomena and design of experiments. By shifting theories from "proposals of truth to be falsified" to "proposed problem-solving tools", the pragmatic view prompts us to assess a theory by its utility: what empirical problems it can solve, how easily it can be used to solve them, and how good its solutions are. It also requires criteria to evaluate the quality of solutions to a problem and a set of standards by which we measure the utility of the theory such as accuracy, simplicity, falsifiability, generalizability, and reproducibility (Chang 2007b, 2011; van Fraassen 1980; Laudan 1978; Schindler 2018). Through competition to solve empirical problems, theories become more precise, provide clearer and more concise explanations, can be used to make more reliable and accurate predictions, and can be applied to larger domains.

*Conceptual frameworks provide constructs and constraints*

Assessing scientific explanations inevitably involves considerations that are not directly related to solution quality, but are instead constraints on the form solutions can take. These constraints constitute a *conceptual framework* (Table 1): a language within which explanations are proposed. In effect, a conceptual framework is a set of foundational theories that provide a conceptual structure on which further theories within that program are built (Lakatos 1978; Laudan 1978; Kuhn 2012).

The stability of such a framework allows its component theories to change without rebuilding their conceptual foundations. For example, under the modern framework of neuropsychiatry, psychiatric disorders are framed in terms of biophysical dysfunctions in neural structure. Current debates about the underpinnings of schizophrenia include hypotheses of dysfunction within dopaminergic or glutamatergic systems, dysfunctional pruning of dendrites, and dysfunctional oscillatory dynamics (Moghaddam and Javitt 2012; Glausier and Lewis 2013; Uhlhaas and Singer 2015; Howes et al. 2017). However, they all lie within a general framework of biophysical changes in neural processes. The consistency of this founding idea allows us to modify theories without disrupting the foundational premise, which allows them to be directly compared and contrasted.

While explanations are naturally comparable within a framework, theories under different frameworks are composed of fundamentally different objects and describe the world in different terms, which makes them difficult to compare. For example, explanations under the traditional psychoanalytic framework (Luyten et al. 2015) are fundamentally different from those under the modern neuropsychiatry framework (World Health Organization 2021; American Psychiatric Association 2013; Cuthbert and Insel 2013; Insel and Cuthbert 2015). The two frameworks are composed of fundamentally different objects and are described in different terms: in contrast to the neuropsychiatric framework, psychoanalytic explanations for schizophrenia invoke unconscious conflicts and/or distorted ego functions as the key factors underlying psychosis (Luyten et al. 2015). Even the categorizations of psychiatric phenomena are different under these frameworks, making direct comparisons of explanations for the same phenomena across frameworks difficult (Feyerabend 1993).

Despite the difficulties in directly comparing theories across frameworks, all frameworks are not equivalent. One can compare conceptual frameworks by asking how well their theories allow us to predict and control our environment (Lakatos 1978). This is not to say that all research requires a direct application, but rather that consideration of practical components is necessary for a complete understanding of scientific progress (Laudan 1978; Douglas 2014). For example, the psychoanalytic framework implies treatment based analytic therapy, while the modern neuropsychiatry framework suggests medication as a key component. Furthermore, under the new framework known as computational psychiatry, psychiatric disorders are attributed to computational “vulnerabilities” in the systems architecture of the brain (A. D. Redish 2004; A. D. Redish, Jensen, and Johnson 2008; Montague et al. 2012; A. D. Redish and Gordon 2016; Huys, Maia, and Frank 2016). Theories in this new framework suggest that such disorders would be treatable by changing information processing – by modifying the physical substrate (e.g. through electrical stimulation or pharmacological changes), enhancing compensation processes (e.g. through cognitive training), or changing the environment (e.g. by giving a student with ADHD extra time on a test). The pragmatic view suggests that the ultimate adoption (or not) of this framework will come down to how successfully it can be applied to unsolved problems.

*Models as the interface between theories and phenomena*

While “theoretical" work may appear further from “applied” science than its experimental counterpart, models can act as an interface between theory and phenomena. A model consists of a structure and an interpretation of how that structure relates to its target phenomena ((Frigg and Hartmann 2006), also known as the model's “construal” (Weisberg 2013)). For example, the equation $\tau dV/dt = -V + V_{rest}$ is a mathematical structure that is interpreted to represent the temporal dynamics of the membrane potential, $V$, of a passive cell with time constant, $\tau$, and resting potential, $V_{rest}$ (Hille 2001; Hodgkin and Huxley 1952; Koch and Segev 1989; Rall 1992; Gerstner et al. 2014). Models whose structure consists of mathematical equations or computational processes are amenable to simulation and analytical treatment. Models can be constructed from many different kinds of interpreted structures, such as physical structures that are interpreted to represent the double helix of DNA (Watson and Crick 1953) or diagrammatic structures that are interpreted to represent protein interactions involved in signaling cascades (Alon 2006). Many “animal models” used in experimental neuroscience are physical structures interpreted to represent other phenomena, such as the 6-OHDA rat or the MPTP monkey, which are interpreted to represent the pathology of Parkinson’s disease (Dorval and Grill 2014; Schultz et al. 1989).

In creating a model, a researcher has to make foundational assumptions in the terms they use, the form those terms take, and the relationships between them. These assumptions instantiate aspects of a theory in an explicit expression with a well-defined form. The voltage equation above instantiates the theory that a neuron’s electrical properties arise from a semipermeable membrane (Hodgkin and Huxley 1952; Rall 1992), while the 6-OHDA model instantiates the theory that Parkinson's disease arises from dopaminergic dysfunction (Langston and Palfreman 2013). This explicit formulation of theories can force us to confront hidden assumptions (Marder 2000), and provide useful insights for the design of experiments or potential engineering applications.

Further, in selecting some aspects of a phenomenon to include, and others to ignore, creating a model abstracts a multi-faceted phenomenon into a concise, but inevitably simplified, representation. Thus, models simultaneously act as an instantiation of a theory and an abstraction of a phenomenon (Rosenblueth and Wiener 1945; Stafford 2009). This dual role of models is the foundation of their use in explanation (Cartwright 1997).

[TABLE 1 NEAR HERE]

## Descriptive, mechanistic, and normative explanation

The terms "descriptive", "mechanistic", and "normative" are widely used in neuroscience to describe various models. A pragmatic view prompts us to consider how these terms relate to the type of problem they are used to solve (Kording et al. 2020). In doing so, we find that these labels correspond to three different explanatory approaches in neuroscience, which are used to solve three different types of problems: "what" problems, "how" problems, and "why" problems (Dayan and Abbott 2001). (See Figure 1).

[FIGURE 1 NEAR HERE]

*Descriptive explanations*

The first problem often encountered in scientific research is: *What is the phenomenon?* Phenomena are not divided into discrete entities *a priori*, but instead appear as a continuous multifaceted stream with many possible methods of observation and many aspects that could be observed. Thus, the set of characteristics that define a phenomenon are often unclear. This problem is addressed with a **descriptive** explanation (David M. Kaplan and Bechtel 2011). For example, to explain the spikes observed from a hippocampal neuron we could use a theory of "place cells" (O'Keefe and Nadel 1978): a collection of ideas that defines the relationship between neural activity in the hippocampus and an animal's position in an environment, which can be instantiated in a model that specifies that relationship in an equation (O'Keefe and Nadel 1978; A. D. Redish 1999; Laura Lee Colgin 2020). Descriptive models are founded on basic assumptions of which variables to observe and how to relate them. At its heart, a descriptive explanation is simply a selective account of phenomenological data; indeed descriptive models are often called phenomenological models (Craver 2007; David Michael Kaplan 2011) or, when they are well-established, phenomenological laws (Cartwright 1997).

*Mechanistic explanations*

After addressing the "what" question, one might ask: *How does the phenomenon arise?* This problem is addressed with a **mechanistic** explanation, which explains a phenomenon in terms of its component parts and their interactions (Machamer, Darden, and Craver 2000; Craver 2007; Bechtel and Richardson 2010). For example, to explain the activity of place cells, we can create explanations based on afferent information from other structures, internal connectivity patterns, and intra-neuronal processing, which can be instantiated in a model that specifies how they interact to produce neural firing (A. D. Redish 1999;

Hartley et al. 2000; Barry et al. 2006; Solstad, Moser, and Einevoll 2006; Fuhs and Touretzky 2006; Giocomo, Moser, and Moser 2011; Sanders et al. 2015). A mechanistic model is founded on an assumption of which parts and processes are relevant, and illustrates how their interaction can produce a phenomenon or, equivalently, how the phenomenon can emerge from these parts. Often these parts are considered to be causally relevant to the phenomenon, and a mechanistic explanation is often also referred to as a causal explanation (Machamer, Darden, and Craver 2000; Craver 2007; Bechtel and Richardson 2010).

Mathematical mechanistic models in neuroscience often take the form of a dynamical system (Koch and Segev 1989; Ellner and Guckenheimer 2006; Izhikevich 2007; Ermentrout and Terman 2010; Gabbiani and Cox 2017; Gerstner et al. 2014; Börgers 2017), in which a set of variables represent the temporal evolution of component processes or their equilibrium conditions. For example, the classic Hodgkin-Huxley model uses a set of four coupled differential equations to represent the dynamics of membrane potential and voltage-dependent conductances, and shows how an action potential can emerge from their interaction by producing a precise prediction of the progression of the membrane potential in time (Hodgkin and Huxley 1952). However, qualitative mechanistic models, in which complex processes are summarized in schematic or conceptual structures that represent general properties of components and their interactions, are also commonly used in neuroscience. For example, Hebb considered a conceptualization of neural processing in which coincident firing of synaptically connected neurons strengthened the coupling between them. From this model, Hebb was able to propose how memories could be retrieved by the completion of partial patterns and how these processes could emerge from synaptic plasticity, as cells that were coactive during a particular stimulus or event would form assemblies with the ability to complete partially-activated patterns (Hebb 1949).

Mechanistic models represent the (assumed) underlying processes that produce the phenomenon (Craver 2007; David M. Kaplan and Bechtel 2011). They can be used to make predictions about situations where the same processes are presumed to operate (Ellner and Guckenheimer 2006). This includes the effects of manipulations to component parts, and circumstances beyond the scope of data used to calibrate the model.

*Normative explanations*

In addition to the mechanistic question of “how”, we can also ask the question: *Why does the phenomenon exist?* This kind of problem is addressed with a **normative** explanation**,** which is used to explain a phenomenon in terms of its function (Barlow 1961; Kording, Tenenbaum, and Shadmehr 2007; Bialek 2012). A normative explanation of place cells would appeal to an animal’s need to accurately encode its location, and could instantiate that need in a model of a navigation task (O’Keefe and Nadel 1978; A. D. Redish 1999; McNaughton and Nadel 1990; Zilli and Hasselmo 2008). Appealing to a system’s function serves as a guiding concept that can be a powerful heuristic to explain its behavior based on what it ought to do to perform its function (Dennett 1989). This kind of explanation has a long history in the form of teleological explanation, which explains a thing by its “purpose” (Aristotle, n.d.), and is often used implicitly in biological sciences — for example, stating that the visual system is “for” processing visual information. In neuroscience, functions often come in the form of cognitive, computational, or behavioral goals.

When quantified, normative models formalize the goal of the phenomenon in an objective function (also known as a utility or cost function), which defines what it means for a system to perform “well”. These models are founded on an assumed statement of a goal and the constraints under which the system operates. For example, models of retinal function formalize the goal of visual processing using equations that represent the ability to reconstruct a sensory signal from neural responses, under the constraints of sensory degradation and a limited number of noisy neurons (Rieke et al. 1997; Field and Rieke 2002; Doi and Lewicki 2014). Such an approach also relies on an assumption of an underlying optimization process. This assumption is often justified by appealing to evolution, which might be expected to optimize systems (Parker and Smith 1990; Barlow 1961; Bialek and Setayeshgar 2008). However, evolution does not guarantee optimality due to limitations of genetic search (Gould 1983; Gould and Lewontin 1979). Moreover, there are numerous processes in physical, biological, neurological, and social systems that can drive phenomena towards a state that maximizes or minimizes some objective function; however, these processes each also have their own unique limitations. For example, physical processes that minimize surface-to-volume ratio create hexagonal tessellations in beehives, but this process is limited by the physical properties of construction (Thompson 1992; Smith, Napp, and Petersen 2021). Economic markets might be expected to optimize the balance between offer and selling price, but are limited by imperfect and unbalanced information and the limited decision-making abilities of agents (Kahneman, Knetsch, and Thaler 1991; Gigerenzer and Gaissmaier 2011; Fox 2009; Shleifer 2000; Akerlof 1978). Similarly, supervised learning might be expected to optimize object discrimination, but its implementation in the brain would be limited by constraints such as synaptic locality and the availability of credit signals and training data (Hunt et al. 2021; Hamrick et al. 2020; Häusser and Mel 2003; Richards et al. 2019; Takeuchi, Duszkiewicz, and Morris 2014; McNaughton, Douglas, and Goddard 1978). Where each of these processes might be expected to bring systems toward an optimal solution, the constraints under which they operate may themselves impose distinct signatures on the systems they optimize.

*The descriptive / mechanistic / normative classification depends on context*

Theories and models do not exist in isolation, but are embedded in scientific practice. As the descriptive/mechanistic/normative categorization reflects the problem being solved, it can be applied to both theories and models depending on the context, i.e., kind of explanation, in which they are being used. In general, this categorization is independent of whether an explanation is accepted by the scientific community. For instance, a mechanistic explanation does not cease to be mechanistic if it is not adopted, e.g., because some of its predictions are not experimentally corroborated. Further, models with the same structure can be used for different purposes, and can thus be assigned to a different category in different contexts. For example, the integrate-and-fire model can be used as a descriptive model for membrane potential dynamics, or as a mechanistic model for the neuronal input-output transformation; and while the Hodgkin-Huxley model was discussed above as a mechanistic model for the problem of spike generation, it was originally proposed to be "an empirical *description* of the time course of the changes in permeability to sodium and potassium" (Hodgkin and Huxley 1952). In fact, theories often start as an effort to solve one class of problem, and over time develop aspects to address related problems of different classes — resulting in a theory with descriptive, mechanistic, and normative aspects.

## Levels of abstraction

In selecting some aspects of a phenomenon to include, and others to ignore, a model abstracts a multi-faceted phenomenon into a more concise, but inevitably simplified, representation. That is, in making a model we replace a part of the universe with a simpler structure with arguably similar properties (Rosenblueth and Wiener 1945; Weisberg 2013). It could be argued that abstraction is detrimental to model accuracy (i.e. that "The best material model for a cat is another, or preferably the same cat"), and is only necessary in light of practical and cognitive limitations (Rosenblueth and Wiener 1945). However, abstraction is important in scientific practice, and its role extends beyond addressing those limitations (Potochnik 2017).

*Descriptive models define abstractions at different levels*

Abstraction is most obvious when we consider the construction of descriptive explanations. First, abstractions are made when researchers decide which aspects of a phenomenon *not* to include. For example, the cable equation which describes the relationship between axonal conductance and membrane potential (Rinzel and Ermentrout 1989; Rall 1992; Gerstner et al. 2014) does not include details about intracellular organelles, the dynamics of individual ion channels, or the impact of nearby neurons on the extracellular potential. Importantly, these models do not include many larger scale effects (such as the neuron's embedding in a circuit, or the social dynamics of the agent) as well as smaller scale factors (Vinogradov, Hamid, and Redish 2022). The process of abstraction thus applies to both phenomena at smaller scales (organelles) and at larger scales (social interactions of the agent) that are hypothesized to be unnecessary to address the question at hand. Each of these factors are abstracted away, leaving only the features chosen to be represented in a model's structure.

Second, the aspects that are included must be represented in an idealized form. For instance, ionic flux through the cell membrane is not a strictly linear current function of voltage and conductance, but we often idealize it as such for tractability (Rall 1992; Koch and Segev 1989; Hille 2001). These idealizations are assumptions about a phenomenon which are, strictly speaking, false, but are used because they serve some purpose in creating the model (Potochnik 2017).

Classic accounts of neuroscience emphasize analysis at different levels of abstraction (Churchland and Sejnowski 1994; Craver 2007; Marr 1982; Shepherd 1994; Sejnowski, Koch, and Churchland 1988; Wimsatt 1976) (Box 1). However, despite the ubiquity of level-based views of neuroscience and a number of proposed schemes, no consensus can be found on what the relevant levels of abstraction are, or even what defines a level (Guttinger and Love 2019). Suggestions of different level schemes range from those of computational abstraction (Colburn and Shute 2007; Wing 2008), which simplifies a process to be independent of its specific implementation or physical substrate, to levels of conceptual abstraction, which delineate the degree of idealization vs relatability to data (O'Leary, Sutton, and Marder 2015), and levels of physical abstraction, which are used to deal with different spatiotemporal scales (Churchland and Sejnowski 1994). However, recent analyses suggest that natural phenomena are not organized into levels in a universally coherent manner (Potochnik and McGill 2012; Potochnik 2017, 2020). From a pragmatic view, levels of abstraction need not reflect discrete "levels" in nature, but are indicative of our problem-solving strategies and constraints. Because different abstractions can facilitate different research aims (Potochnik 2017), multiple descriptive models are needed to represent the same phenomenon that abstract different features to different degrees.

[BOX 1 NEAR HERE]

*Mechanistic and normative models connect levels of abstraction*

Without links between them, we would be left with a hodgepodge of different descriptions. However, unification has been noted as a strong desideratum for scientific theories (Schindler 2018; Keas 2018). The relationship between different descriptions of the same phenomena can often be expressed in terms of a **mechanistic explanation**. For example, we might describe single-neuron activity in terms of membrane currents, or by listing a set of spike times: a natural reduction in the dimensionality that can result from many possible combinations of currents (Golowasch et al. 2002; Prinz, Bucher, and Marder 2004). A mechanistic model (e.g., (Hodgkin and Huxley 1952)) that demonstrates how spike times emerge from currents connects the descriptions at the two levels and, in addition, does so asymmetrically, as it does not claim to be a mechanism by which currents emerge from spike times. By bridging descriptions that each abstract different features to different degrees, mechanistic explanations create a multi-level 'mosaic unity' in neuroscience (Craver 2007), in which descriptions are grounded through their interconnections, and more abstract features are grounded in their emergence from less abstract counterparts (Craver 2007; David M. Kaplan and Bechtel 2011; Bechtel 2008; Craver 2002; Oppenheim and Putnam 1958).

In contrast, a **normative explanation** connects descriptions by appealing to the ability of less abstract features to satisfy a description of more abstract goals. For example, the mammalian hypothalamus could be described as maintaining body temperature like a thermostat (Tan and Knight 2018; Morrison and Nakamura 2011) or as a circuit of interconnected neurons. A normative model connects the two descriptions by explaining the negative feedback loop in the circuit through its ability to achieve those thermostatic functions. Because functions exist over a range of levels, from cellular to behavioral or computational, we could imagine a “multi-level” approach to understanding the mammalian hypothalamus that in turn uses the goal of a negative feedback loop to explain the developmental processes that establish hypothalamic connectivity. Like their mechanistic counterparts, normative explanations establish links between descriptions which each have their own utility for different problems, by virtue of their unique abstractions.

Thus, the three-fold division of explanatory labor in neuroscience falls naturally into the different roles a model can play in terms of levels of abstraction. **Descriptive explanations** define abstractions of phenomena at different levels, while **mechanistic** and **normative explanations** bridge levels of abstraction. Descriptive models, rather than “mere” descriptions of phenomena (as they’re sometimes dismissed), are the necessary foundation of both normative and mechanistic models. In turn, mechanistic and normative explanations connect a description at a “source” level to a description at a higher or lower “target” level (Figure 2). Each of the terms that represent the components of mechanistic models and the constraints of normative models are descriptive models at a lower level of abstraction, while those that represent the emergent properties of mechanistic models and the goals of normative models are descriptive models at a higher level of abstraction. Given their multi-level nature, a dialogue between descriptive, normative, and mechanistic models is needed for a theoretical account of any neuroscientific phenomenon.

[FIGURE 2 NEAR HERE]

*At what level of abstraction should a model be built?*

As different abstractions trade-off advantages and disadvantages, the selection of which abstraction to use is highly dependent on the problem at hand (Herz et al. 2006). Current neuroscientific practice generally attempts two approaches for selecting the appropriate level of abstraction, which serve different purposes. The first approach is to try to find as low a level as possible that still includes experimentally-supported details and accounts for the phenomenon. For example, one might explain the phenomenon of associative memories using compartmental models of pyramidal cell networks, including specific active conductances, dendritic compartments, pharmacological effects on different inputs arriving at different compartments and identifying the consequences for learning and recall (Hasselmo 1993). The multiplicity of parameters and variables used in this approach provides many details that can be matched to observable features of a phenomenon and can capture unexpected properties that emerge from their interaction. However, these details need to be extensively calibrated to ensure the model is accurate, and can be very sensitive to missing, degenerate, or improperly tuned parameters (Traub, Jefferys, and Whittington 1999; Traub et al. 1991). The second approach is to try to find the most abstract level that can still account for the phenomenon. For example, we might instead appeal to the classic Hopfield network, in which units are binary (+1, -1), connections are symmetrical, and are updated using a very simple asynchronous rule (Hopfield 1982; Hertz, Krogh, and Palmer 1991). While more abstract models sacrifice the ability to make predictions about lower level details, their insights are often more robust to specific (e.g. unobserved) physiological details, and by reducing a complicated system to a small number of effective parameters, they allow for powerful analysis on the influences to the system properties. Further, abstract models can provide conceptual benefits such as intuition for how the system works and the ability to generalize to other systems that can be similarly abstracted (Gilead, Trope, and Liberman 2019; Gilead, Liberman, and Maril 2012; O'Leary, Sutton, and Marder 2015).

Another important consideration is the ability of models at different levels to interface with different experimental modalities or scientific fields. Every measurement is itself an abstraction, in that it is a reduced description of the part of the universe corresponding to the measurement (Chang 2007b). For example, fMRI measures blood flow across wide swaths of cortex, but abstracts away the interactions between individual neurons, while silicon probes measure extracellular voltage but abstract away intracellular processes, and calcium imaging measures neuronal calcium levels, but abstracts away the electrophysiology of neuronal spiking. All of these are discussed as “neural activity”, but they likely reflect different aspects of learning, performance, and dynamics. Moreover, subsequent processing abstracts these signals even further, such as correlation (functional connectomics) in fMRI, sorting voltage signals into putative cell “spiking” from silicon probes, and treating calcium transients as “events” from calcium imaging. The abstraction made by one measurement device might lend itself to explanations at a given level, but not others, and the measurements available are important considerations when selecting which abstractions to make in our models.

Similarly, models at different levels are often used by distinct scientific fields or communities. The existence of a literature with a rich body of relevant work can provide details and support for components of a model outside of the immediate problem of interest. Integrating theories and models across these different fields can be particularly beneficial for scientific progress (Wu, Wang, and Evans 2019; Grim et al. 2013); however, crossing levels can be a sociological problem as well as a methodological one because different fields of study often use different languages and operate under different conceptual frameworks.

In general, it is important that researchers spell out the abstractions being made in their models, including their purposes as well as their limitations. By being concrete about the abstractions made, researchers can increase the reliability of their theories. Importantly, as noted above, it is useful to acknowledge not only the simplifications made about smaller-scale phenomena, but also the simplifications made as to larger-scale interactions that have been abstracted away from a theory.

## Theory development and experimentation

Traditional views emphasize the use of experiments to test proposed theories (Popper 1959), and even consider an interplay in which theories suggest new experiments and unexpected experimental results reveal the need for new theories (Firestein 2015; Laudan 1978). However, theories do not arise fully-formed, but are developed over time through an interaction with experimentation (Bechtel 2013; Laudan 1978; Hacking 1983; Douglas 2014; Firestein 2015). We now consider two crucial pieces of that dialogue: the **domain** of a theory, or phenomena it is intended to pertain to, and a **translation function**, which specifies how it should relate to phenomena in its domain. Experimentation plays two key roles in relation to theory: 1) grounding model assumptions and 2) assessing the quality of model-based explanations. We then discuss an often underappreciated form of experimentation, in which models themselves are the experimental subjects. These modeling experiments allow us to explore the (sometimes hidden or unexpected) implications of a theory itself, identify its underlying inconsistencies, and can be used to predict novel phenomena. Together, this reveals a picture in which theory development is not relegated to simply proposing theories-to-be-tested, but instead entails a complex experimental paradigm in which models play an active role in the simultaneous development, assessment, and utilization of theories within explicit conceptual frameworks.

*Linking Theory and Phenomena*

The **domain** of a theory is the set of phenomena that it purports to explain (Kuhn 2011, 2012; Mitchell, Keller, and Kedar-Cabelli 1986; A. D. Redish 1997). The domain is therefore a set of data-imposed constraints, and the theory should provide an explanation consistent with those constraints. Theoretical studies should be explicit about what phenomena do and do not lie in their intended domain. In practice, nascent theories are often evaluated not only by their ability to explain data in their proposed domain (Feyerabend 1993; Laudan 1978) but also by their potential to expand beyond the initial domain with further development (Lakatos 1978). For example, the theory that action potentials arise from voltage-dependent changes in ionic permeability (Hodgkin and Huxley 1952; Goldman and Morad 1977; Katz 1993; Hille 2001) should apply to the domain of all action potentials in all neurons. Early theories of action potential function identified voltage-gated sodium currents as the primary depolarizing component and formalized their action in models that developed into the Hodgkin-Huxley framework (Hodgkin and Huxley 1952). When some action potentials were later found to be independent of sodium concentrations, it was straightforward to incorporate other voltage-gated channels within the same framework (Hille 2001; Koch and Segev 1989; Gerstner et al. 2014).

By instantiating a theory in a specific structure (Rosenblueth and Wiener 1945; Stafford 2009), models play a key role in connecting a theory to phenomena in its domain. However, no model is directly comparable to experimental data by virtue of its structure alone. As noted above, a model also consists of an interpretation of how that structure relates to its target phenomena (Weisberg 2013). This interpretation is specified by a **translation function**: a statement of how the model's components map onto its target phenomena. A translation function may be as straightforward as "variable *V* represents the membrane potential in millivolts", but it can also be less constrained, e.g., "variable *V* describes the slow changes in the membrane potential and ignores all spiking activity". In other cases, the translation function can be complex, as parts of the model can have a loose correspondence to general features of large classes of data, and can represent highly abstract effective parameters or qualitative behaviors. For example, the units in Hopfield's attractor network models (Hopfield 1982; Hopfield and Tank 1985; Hertz, Krogh, and Palmer 1991) are not meant to directly correspond to measurable properties of biological neurons, but are instead intended to reflect qualitative features, namely that neural populations are "active" or not. In effect, the translation function spells out the abstractions made by the model. Specifying the translation function of a model is as important as defining its structure (Weisberg 2013). While these descriptions are often provided for highly abstract models, models that describe finer spatio-temporal scales (such as detailed compartmental models of neurons) are often considered to be "biologically realistic" and assume a simple or obvious translation function. However, it is important to remember that these models are also abstractions, albeit at a different level, and a proper description of the abstractions made will help clarify both the uses and the limitations of such models. By specifying the intended correspondence between model terms and phenomena, the translation function operationalizes the concepts associated with those terms in the theory (Bridgman 1927; Chang 2007a).

[FIGURE 3 NEAR HERE]

*Experiments ground model assumptions*

With a well-defined translation function in hand, we can now consider the ways in which models are informed by experimental data. As outlined above, the components of descriptive, mechanistic, and normative models are each based on a different set of foundational assumptions. These assumptions are not generally arbitrary, but are informed by experimental observations and results.

**Descriptive models** are founded on an assumed relationship between variables, which is generally formulated to capture an observed regularity in experimental data. These initial observations often rely on "exploratory" experiments, which attempt to identify empirical regularities and the constructs with which to describe them (Steinle 1997). In specifying the characteristic properties of a phenomenon, descriptive explanations delineate the attributes that are expected to be replicable in future experiments and play a foundational role in subsequent mechanistic and normative models. This is extremely important for the current replication controversy (Baker 2016; Goodman, Fanelli, and Ioannidis 2016; Fanelli 2018; A. D. Redish et al. 2018). A recent National Academy report (National Academies of Sciences and Medicine 2019) characterizes *replicability* as the ability to obtain consistent results across multiple studies, and contrasts it with *reproducibility*, defined as the ability to get the same results when applying the same analyses to the same data. Several authors have suggested that the replication crisis is in fact a crisis of theory development, as it is the scientific claims (not data) that should be replicable (drugmonkey 2018; A. D. Redish et al. 2018; Smaldino 2019). We suggest that this crisis stems from three sources: 1) a failure to define domains correctly, assuming that limited observations correspond to a much larger range of phenomena than they actually do, 2) a failure to formalize observations in adequate descriptive models (e.g., an overreliance on correlation, or assumed simple relationships), and 3) a failure to connect those descriptive models with mechanistic or normative models that integrate descriptions at different levels of abstraction.

**Mechanistic models** are founded on a set of parts and interactions that are assumed to be relevant to a target phenomenon. The existence of candidate parts/interactions can be informed by experimental observations, and their relevance (or irrelevance) to a given target phenomenon is often derived from experimental or natural interventions (Pearl 2009). Once the decision is made to include a part/interaction in a mechanistic model, its corresponding terms can be parameterized by virtue of the descriptive models at their source level of abstraction. For example, when trying to explain the phenomenon of burst spiking in thalamocortical neurons, we might observe the presence of a hyperpolarization-activated current ($I_h$) which, when blocked, disrupts burst spiking (McCormick and Pape 1990). We can then calibrate the parameters used to model $I_h$ with values acquired through slice experiments.

Experimental data can also inform the founding assumptions (goal/constraints) of **normative models**. For example, when trying to explain the responses of visual neurons, we might parameterize the constraints of an efficient coding model with data from retinal photoreceptors (Field and Rieke 2002). As with mechanistic models, normative parameters rely on the descriptive models we have for photoreceptor properties. However, grounding an assumed function (e.g. “vision”) in experimental data can be more challenging. This arises from a notable asymmetry between mechanistic and normative approaches: while the founding assumptions of a mechanistic model (parts/interactions) are *less* abstract than their target phenomena, the founding assumptions of normative approaches (a function/goal) are generally *more* abstract than the phenomenon they are used to explain. This often results in normative approaches being termed “top-down”, in contrast to “bottom-up” mechanistic modeling. In practice, functions are often operationalized via performance on a specified task, rendering them groundable in experimental data. For example, the assumed goal of primate facial recognition areas is grounded in the change in facial recognition abilities when those neural systems are manipulated or absent, neural responses to facial stimuli, and in the coupling of those areas with sensory and motor areas providing a behavioral circuit (Gross, Bender, and Rocha-Miranda 1969; Tsao et al. 2006; Moeller et al. 2017; Grimaldi, Saleem, and Tsao 2016).

*Experiments assess solution quality*

As has been noted by many previous authors, we cannot definitively “confirm” theories (Popper 1959), nor can we definitively test/falsify the validity of a theory in isolation (Duhem 1991; Lakatos 1980). However, a theory’s utility does not require absolute confidence in its validity, but only a track record of solving problems in its domain. By instantiating theories in a model with a well-defined translation function, we can assess the quality of solutions proposed with a given theory by comparing the behavior of those models to experimental observations.

In the case of **descriptive models**, model fitting can estimate confidence intervals and goodness-of-fit for the best-fitting parameter values, and can even be used to quantitatively compare candidate models to determine which can best explain experimental data with the fewest parameters. A researcher might build a **mechanistic model** with terms that correspond to the proposed parts to see if they are able to reproduce features of the data, or test the model's ability to predict the effect of experimental manipulations. Alternatively, a researcher can hypothesize that the system is performing some function, make a **normative model** that instantiates the goal, and see if properties of the data match those expected from a system optimizing that goal. In these "confirmatory" (theory-driven/hypothesis-testing) experiments, models are used to apply existing theories to account for observed phenomena, compare possible instantiations of a theory, or even compare theories with overlapping domains to see which better accounts for the phenomenon. In each case, the assumptions of the model act as a hypothesis and the degree of similarity between model and experimental data is used to assess the *sufficiency* of a theory (and its specific model instantiation) to account for a phenomenon.

However, the value of modeling is often in its ability to show *insufficiency* of a theory/model to account for experimental data. Rather than invalidating the theory, this can often prompt updates to the theory or a search for yet-unobserved relevant phenomena. For example, early models of head-direction tuning found that a mechanism based on attractor networks required recurrent connections not supported by anatomical data (A. D. Redish, Elga, and Touretzky 1996). This incompatibility led to subsequent analyses which found that the tuning curves were more complicated than originally described, matching those seen in the model without the recurrent connections (Blair, Lipscomb, and Sharp 1997). Similarly, the usefulness of normative models often lies in their ability to identify when a system is performing suboptimally (Parker and Smith 1990). Such a finding can provide additional information about unexpected functions or constraints. When there is a mismatch between a normative model and observed phenomena, one could hypothesize that the agent is optimizing a different goal (Fehr and Schmidt 1999; Binmore 2005), new constraints that limit the processes available (Simon 1972; Mullainathan 2002), historical processes that could limit the optimization itself (Gould and Lewontin 1979; Gould 1983), or computational processes that limit the calculations available to the system (A. David Redish 2013; Nadel 1994; Schacter 2001; Webb, Glimcher, and Louie 2021). For instance, several studies have found that foraging subjects tend to remain at reward sites longer than needed (Nonacs 2001; Camerer 1997; Carter and Redish 2016) and accept longer-delay offers than would be expected to maximize total reward (Wikenheiser, Stephens, and Redish 2013; Sweis et al. 2018; Schmidt, Duin, and Redish 2019; Sweis, Thomas, and Redish 2018). However, optimality could be restored by assuming an additional factor in the cost function (Simon 1972) subsequently characterized as "regret": an increased cost of making a mistake of one's own agency compared to equivalently poor outcomes that were not due to recognizable mistakes (Wikenheiser, Stephens, and Redish 2013; Steiner and Redish 2014; Sweis, Thomas, and Redish 2018; Zeelenberg et al. 2000; Coricelli et al. 2005). Similarly, Fehr and colleagues have found that normative explanations of behavior in a multi-player game require an additional component with information about one's companion's success in addition to one's own, in order to account for the observed behavior (Fehr and Krajbich 2014; Fehr and Schmidt 1999; Binmore 2005).

*Modeling experiments explore theory implications*

Confirmatory experiments can even be carried out without direct comparison to data, as phenomena at both the target and source levels of abstraction can be pure theoretical entities. Similar to their benchtop counterparts, we can treat different parameters or model instantiations as independent variables in the experiment, and test their sufficiency to account for different aspects of the phenomenon as the dependent variables (Omar, Aldrich, and Gerkin 2014; Gerkin, Jarvis, and Crook 2018). One can use these models as experiments to test the feasibility of theoretical claims in tractable idealized systems. For example, Hopfield's attractor network models (Hopfield 1982; Hopfield and Tank 1985) provided strong support for Hebb's theory (Hebb 1949) that increased connectivity from co-active firing could create associative memory, by showing that strong connections between simple neuron-like entities were sufficient to produce cell assemblies that could be accessed through a pattern-completion process (Hertz, Krogh, and Palmer 1991).

Like their physical analogues (e.g. the 6-OHDA rat or the MPTP monkey), models can be used for exploratory experiments as well. Exploration of the Hopfield model (Hopfield 1982; Hopfield and Tank 1985; Kohonen 1984, 1980) revealed novel properties of categorization, tuning curves, and pattern completion in the neuron-like entities, which were later identified experimentally (K. Obermayer, Blasdel, and Schulten 1992; Klaus Obermayer et al. 2001; Swindale and Bauer 1998; Swindale 2004; de Villers-Sidani and Merzenich 2011; Nahum, Lee, and Merzenich 2013; Freedman et al. 2001, 2003; Lakoff 1990; Rosch 1983; Wills et al. 2005; L. L. Colgin et al. 2010; Yang and Shadlen 2007; Jezek et al. 2011; Kelemen and Fenton 2016). Exploratory modeling experiments can instantiate idealized aspects of a theory to help build intuition for the theory itself. Hopfield's model and its subsequent derivatives have provided researchers with a deeper understanding of how memories can be accessed by content through pattern-completion processes and given rise to concepts such as "basins of attraction" (Hopfield 1982; Hertz, Krogh, and Palmer 1991). These computational discoveries can help build understanding of the theory, and lead to predictions and ideas for new experiments.

Modeling experiments are especially useful in the context of theory development (Guest and Martin 2020). When a phenomenon cannot be readily explained using an existing theory, assumptions can be made as the basis of a modeling experiment. The behavior of this model can then be used to evaluate the sufficiency of these assumptions to account for the phenomenon. Often, these modeling experiments precede a well-formed theory, and a theorist will perform numerous experiments with different models in the process of developing a theory (van Rooij and Baggio 2020). Over time, specific successful model formulations can become closely associated with the theory and develop into its canonical instantiations that make the theory applicable to a wider range of problems and give more precise solutions.

[BOX 2 NEAR HERE]

## Conclusions

A scientific theory is a thinking tool: a set of ideas used to solve specific problems. We can think of theoretical neuroscience as a field which approaches problems in neuroscience with the following problem-solving methodology: **theories** exist within **conceptual frameworks** and are instantiated in **models** which, by virtue of a **translation function**, can be used to assess a theory's ability to account for phenomena in the theory's **domain** or explore its further implications. (See Figure 3.)

We identified three kinds of explanations that play distinct roles in this process: those in which **descriptive theories and models** are used to define the abstractions by which we describe a phenomenon; those in which **mechanistic theories and models** are used to explain phenomena in terms of lower-level parts and their interactions; and those in which **normative theories and models** are used to explain phenomena in terms of a function at a higher level of abstraction.

These considerations lead to a more concrete view of theory in neuroscience under the pragmatic view: a theory is a set of assumptions available to be instantiated in models, whose adequacy for problems in their domain has been vetted via experimentation, and with a well-established translation function that defines their connection to phenomena. Over time and through the development of canonical model formulations, theories become more rigorous, such that researchers agree on how they should be implemented to explain specific domains. A theory in this sense is not a formal set of laws, but a continuously developing body of canonical models and model-phenomenon correspondences, bound together partly by history and partly by shared problem-solving methods and standards (Bechtel 1993).

What recommendations can we take away from this perspective? First and foremost, that scientists should be explicit about the underlying components of their theory. Reliability of theoretical work depends on being explicit about the **domain** that the theory purports to cover, the **abstractions** used (what has been ignored and left out), and the **translation function** to connect the theory to actual measurements. Furthermore, thinking of the pragmatic aspects suggests being explicit about what problems the work proposes to solve, what **conceptual frameworks** the theory fits in, and what the founding assumptions of the **models** are.

Finally, it is interesting to consider that we might apply our taxonomy to our own framework. The concept that 'the ultimate goal of a theory is to provide tools that allow one to better explain and control one's environment' is a normative theory of the goal of scientific theories; the concept that 'models instantiate theories and allow one to test their viability and their relationship to phenomena' is a mechanistic theory of how those theories achieve that goal; and the concept that 'theories live within a framework that a community applies to them' is a descriptive theory of theories. One could imagine a metascientific research program which studies the available phenomena - for example, the scientific literature - to test and further develop those theories, and even the use of models of the scientific process itself (e.g. (Devezer et al. 2019)). The benefits of such a research program could prove as impactful for scientific practice as other theories have proven for manipulation of phenomena in their domain.

## Figure captions

**Figure 1:** The three explanatory processes that underlie scientific explanations. **Descriptive** theories address the question of "what is the phenomenon?" and identify the repeatable characteristics of that phenomenon. **Mechanistic** theories address the question of "how does the phenomenon arise?" and explains the phenomenon in terms of the parts and interactions of other phenomena at lower levels of abstraction. **Normative** theories address the question of "why do the phenomena exist?" and allow a comparison of the phenomenon to an identified function or goal. Normative theories allow the determination of whether a process is achieving its goal — inadequacies generally imply an incomplete understanding of the limitations engendered by processes at a lower level of abstraction.

**Figure 2:** Interactions between three explanatory processes and levels of abstraction. **Descriptive** explanations define an idealized abstraction of specific aspects of a phenomenon for discussion, measurement, and repeatability. **Mechanistic** explanations account for properties of a phenomenon by their emergence from less abstract phenomena, while **normative** explanations account for those properties by appealing to their ability to perform more abstract goals.

**Figure 3**: How the various components discussed in this manuscript interact. The **domain** of a theory is the set of phenomena which it purports to explain. Theories are instantiated in models, which are an abstraction of phenomena in the domain, as specified by a translation function. By constraining the form solutions can take, a **conceptual framework** defines a way of looking at a problem, within which models and theories can be proposed. Note that a given model can instantiate more than one theory and a theory can be instantiated by more than one model.

## Table captions

**Table 1**: Terminology used in this manuscript. Three neuroscience examples.

## Box captions

**Box 1**: Levels of abstraction

**Box 2**: What makes a good neuroscientific theory?  What makes a good model?

*Acknowledgments*

This paper is the result of discussions as part of the workshop “Theoretical and Future Theoretical Frameworks in Neuroscience” (San Antonio, TX, Feb 4-8, 2019) supported by the NSF grants DBI-1820631 (HGR) and IOS-1516648 (FS). The authors acknowledge support from the grants NIH T90DA043219 and The Samuel J. and Joan B. Williamson Fellowship (DL), IRP-NIH ZIA-AA000421 and DDIR Innovation Award, NIH (VA), MH118928 (ZSC), NINDS 1U19NS112953, NIDCD 1R01DC018455, NIMH 1R01MH106674, NIBIB 1R01EB021711 (RCG), NIDCD R01DC014101, Hearing Research Incorporated, Sandler Foundation (AH), H2020 GAMMA-MRI (964644) and H2020 IN-FET (862882) (RBJ), NINDS 1R03NS109923 and NSF/NCS-FO 1835279 (JDM), NIMH-NIBIB BRAIN Theories 1R01EB026939 (FS), IBM Exploratory Research Councils (RT), DOD ARO W911F-15-1-0426 (AA), NSF CRCNS-DMS-1608077, NSF-IOS 2002863 (HGR), NIH MH060605 (FN), NIH MH080318, MH119569, and MH112688 (ADR).

The authors further acknowledge the University of Texas at San Antonio (UTSA) Neuroscience Institute and the New Jersey Institute of Technology (NJIT) Department of Biological Sciences and Institute for Brain and Neuroscience Research for technical support in the organization of the workshop, as well as all of the participants in the workshop. We thank Erich Kummerfeld, Hal Greenwald, Kathryn McClain, Simón(e) Sun, and György Buzsáki for comments on parts of the manuscript, and Matt Chafee and Sophia Vinogradov for help with citations.

## Table 1

Terminology used in this manuscript. Three neuroscience examples.

| | *Examples* | | |
|---|---|---|---|
| | *Cellular* | *Systems* | *Disease* |
| ***Framework*** A general description about the structure of the world, providing a language and a conceptual basis for developing theories. | Explanations for differences in neural functional properties can be appropriately described in terms of differences in the electrochemical properties of membranes and proteins. | Explanations of the production of movement by skeletal muscle contractions can be appropriately described in terms of patterns of action potentials in the central nervous system. | Explanations of neurodegenerative diseases can be appropriately described in terms of dysfunction in cellular processes. |
| ***Theory*** A set of ideas that can be used to explain a set of phenomena (the domain of the theory). | Specific voltage gated ion channels enable excitable properties of neurons such as the action potential. | Many movements are generated by central pattern generators that are primarily driven by internal oscillatory dynamics. | Parkinson's disease is due to loss of dopaminergic function in the substantia nigra. |
| ***Model*** An instantiation of aspects of a theory in an (often mathematical) structure, which is interpreted to represent aspects of a phenomenon. | The Hodgkin-Huxley equations represent the voltage-dependent conductances that underlie the action potential. | Half-center oscillators represent neural circuits in the notochord that underlie swimming processes in the lamprey. | Dopaminergic loss caused by 6-OHDA in rodents and MPTP in non-human primates represent similar losses in Parkinson's disease that underlies behaviors such as bradykinesia and tremors. |

## Box 1: Levels of abstraction

An illustrative example of levels of abstraction comes from computer science (Colburn and Shute 2007; Wing 2008), in which higher-level languages abstract the details specified in lower-level languages by concealing detailed code in a single function that provides the same relationship. Computational abstraction simplifies a process, such that it is independent of its component processes or even its physical substrate. For example, there are many algorithms that sort a list of numbers, but any computational sort command produces the same output regardless of the algorithm used. Computational abstraction is used in neuroscience, for example, when we simplify the molecular process of synaptic transmission in a more abstract model that represents its net effect as an increased firing rate of a postsynaptic neuron. This simplification is akin to conceptual abstraction (O'Leary, Sutton, and Marder 2015), by which more abstract, or idealized, models aim to capture general properties of a process rather than the specific details of any one event or dataset. In neuroscience, computational abstraction is often discussed in terms of David Marr's three levels of analysis (Marr 1982; Pylyshyn 1984): the implementational level is a low-level, concrete statement of a phenomenon, the algorithmic level is an abstraction of the implementational level, explaining the process by which the phenomenon occurs, and the computational level is a high-level (normative) statement of the goal of the process.

Distinct levels of abstraction also arise in neuroscience when considering problems at different spatiotemporal scales (Churchland and Sejnowski 1994). For example, we might consider synaptic transmission in terms of the interactions of various proteins at nanometer to micrometer scales, or we might consider a more abstract model in which neural activity is propagated across the cortex at scales of millimeters or centimeters. When we model phenomena at a given spatiotemporal scale, we make an abstraction that prioritizes organizational details at that scale (e.g., cellular), while further simplifying details at others (e.g., subcellular and network) (Eronen and Brooks 2018). One promising perspective on the emergence of spatiotemporal levels suggests that models at higher levels of abstraction arise from their lower level counterparts via a natural dimensionality reduction of the parameter space (Machta et al. 2013; Transtrum et al. 2015). Such a reduction is possible because models of complex systems are "sloppy": they have a large number of dimensions in parameter space along which model parameters can vary without affecting relevant macroscopic observables (i.e. the microscopic parameters are degenerate with respect to macroscopic behavior (Gutenkunst et al. 2007); for examples in neuroscience, see e.g., (Prinz, Bucher, and Marder 2004; Panas et al. 2015). Thus, abstraction from lower to higher spatio-temporal scales can be seen as a reduction of the lower level parameter space that removes sloppy dimensions, but preserves "stiff" dimensions that have strong influence on observable properties at the higher level. The appropriate dimensionality reduction could be as simple as taking the mean or asymptote of some parameter over a population (Wilson and Cowan 1973; Pinto et al. 1996; Destexhe and Sejnowski 2009), or the set of microscopic parameters needed to produce the same macroscopic behavior might be nonlinear and complex (Prinz, Bucher, and Marder 2004; Transtrum et al. 2015; Jalics, Krupa, and Rotstein 2010; Rotstein et al. 2006)

## Box 2: What makes a good neuroscientific theory?  What makes a good model?

**Be specific.** A theory should be specific, particularly in terms of what the theory is attempting to explain and the strategies for doing so. The theory should define what problems it is trying to solve, and provide the criteria for an adequate solution. It is important to define the *descriptive*, *mechanistic*, and *normative* components of the theory and the rationale behind their selection.

**Identify the domain.** The theory should define the set of questions and problems that it is trying to solve. Importantly, this definition should be a reasonable space of phenomena such that it is easy for someone to determine if a new experiment falls within the domain of the theory or not.

**Specify which aspects of the theory are instantiated in each model, and how.** As shown in Table 1, models instantiate theories, enabling them to be compared with data. It is important to specify which aspects of the theory are instantiated in the model and how those aspects are instanted. It is also important to identify how those aspects were chosen, whether from experimental measurements, theoretical assumptions, or best-fit solutions (or arbitrarily).

**Specify the translation function for all models**. All models require translation to be compared to data. While sometimes those translations will be straight-forward, they usually are not. However, even in situations where the translation is straight-forward, being explicit about the translation function will make clear what data it explains and what experimental predictions it makes.

**Identify the abstractions.** Models at all levels can be useful, but in order to be useful, one must identify what aspects of the world are being abstracted away. It is important to include both abstractions of low-level phenomena and what additional (potentially higher-level) complexities are being ignored.

**Define which aspects of the research are exploratory and which are confirmatory.** The fact that models are a form of experiment creates a way forward for theoretical grant proposals. For example, a researcher can propose to build a model that crosses levels in order to address the question of theoretical viability. Such a proposal may have preliminary data to show that one can build models at each level, even if the researcher has not yet put those levels together. Similarly, a grant proposal can define the domain even if the literature review is incomplete. One can also identify how one is going to explore the parameter space of a set of models to determine how those parameters affect phenomena across levels.

By being explicit about the scientific question being addressed, about the assumptions of the theory, the domain the theory is purporting to address, and the process of building and testing models underlying that theory, grant proposals could be viable even if the theory itself remains incomplete. We call on funding agencies and reviewers to recognize that theory is the foundation of any science, and that construction of rigorous theory and systematic computational modeling are time-consuming processes that require dedicated personnel with extensive training. Our hope is that the framework and associated language outlined in this document can be used to specify deliverables that can be understood by both funders and investigators.

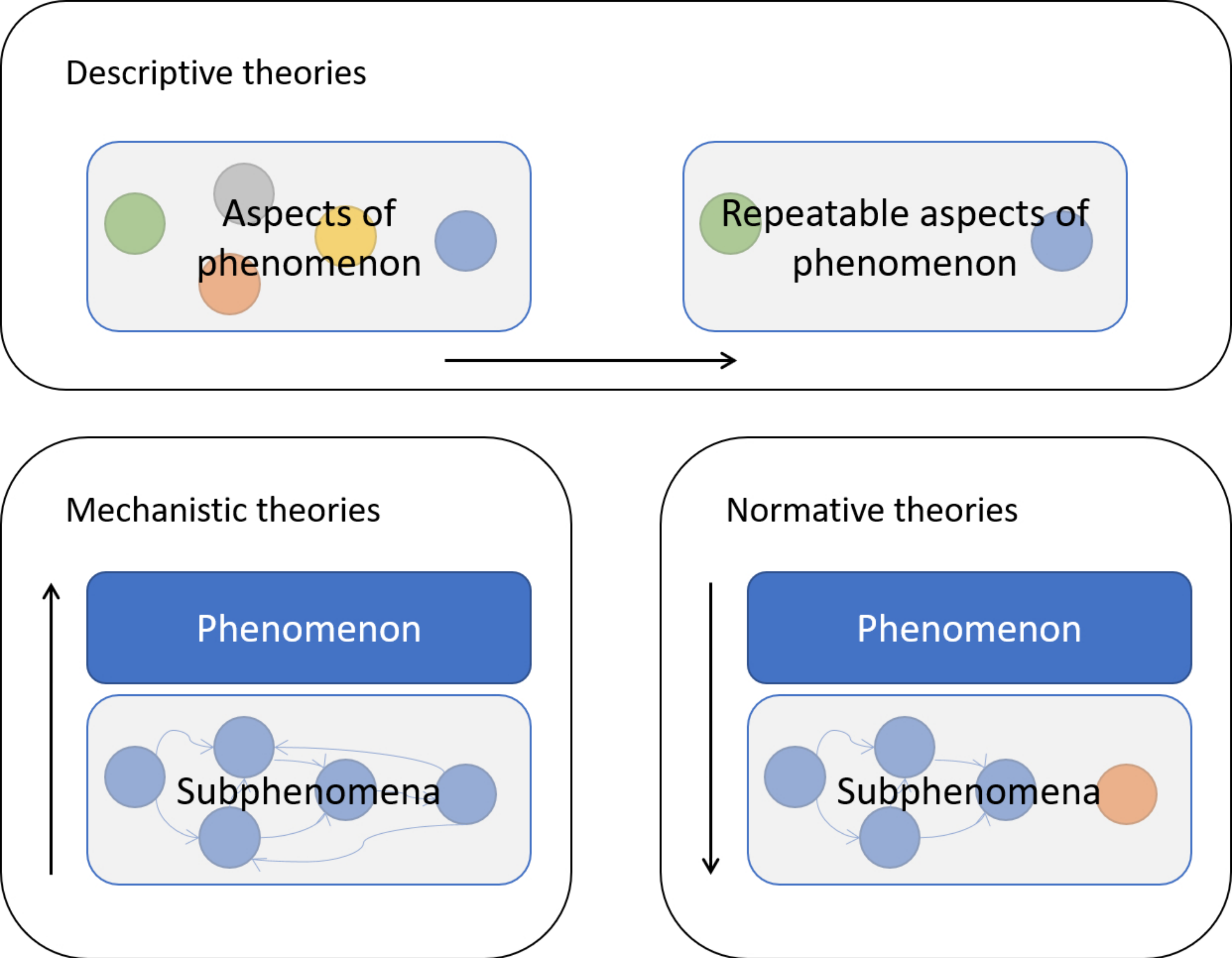

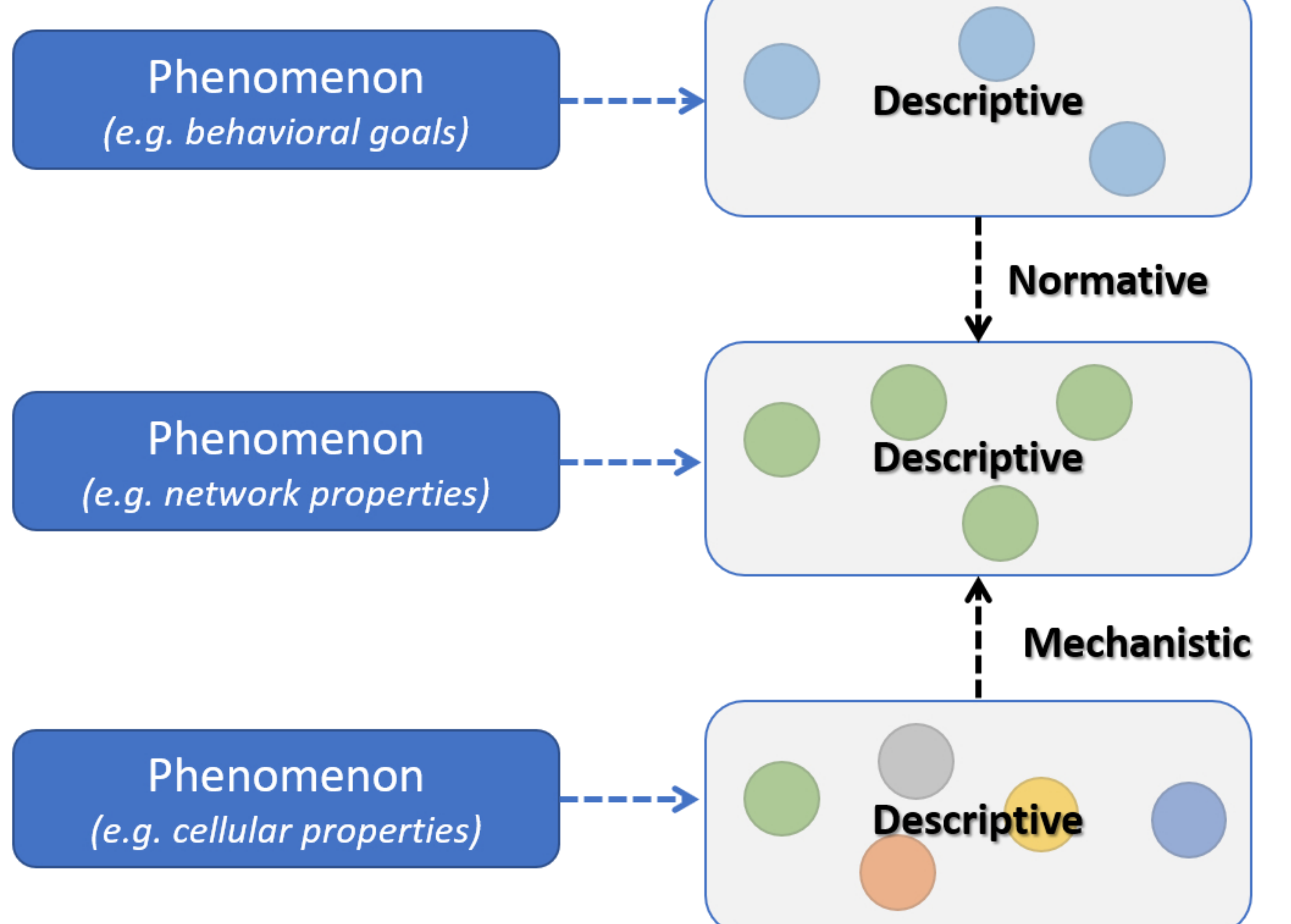
Increasing abstraction (reduced details included)
Phenomenon
(e.g. behavioral goals)
Descriptive
Normative
Phenomenon
(e.g. network properties)
Descriptive
Mechanistic
Phenomenon
(e.g. cellular properties)
Descriptive
(e.g. explain network phenomena by appealing to behavioral goals)
(e.g. explain network phenomena by emergence from cellular properties)

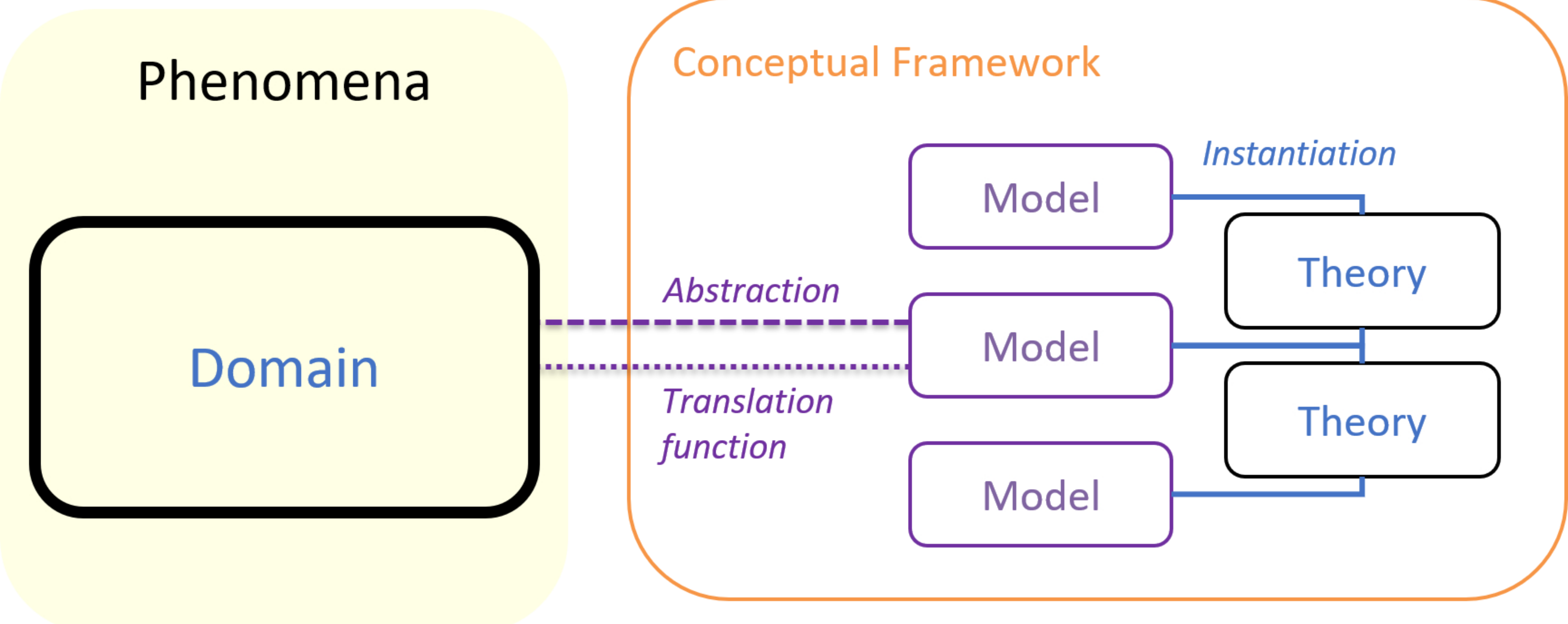